\documentclass[twocolumn,showpacs,preprintnumbers,amsmath,amssymb,prl]{revtex4}

\usepackage[dvips]{graphicx}
\usepackage{dcolumn}
\usepackage{bm}

\begin{document}


\title{
Criterion for weak spin-orbit coupling in heavy-fermion superconductivity:\\
A numerical renormalization-group study
}

\author{Hiroaki Kusunose}
\affiliation{Department of Physics, Tohoku University, Sendai 980-8578, Japan}

\date{\today}

\begin{abstract}
A criterion for effective irrelevancy of the spin-orbit coupling in the heavy-fermion superconductivity is discussed on the basis of the impurity Anderson model with two sets of Kramers doublets.
Using Wilson's numerical renormalization-group method, we demonstrate a formation of the quasiparticle as well as the renormalization of the rotational symmetry-breaking interaction in the lower Kramers doublet (quasispin) space.
A comparison with the quasispin conserving interaction exhibits the effective irrelevancy of the symmetry-breaking interaction for the splitting of two doublets $\Delta$ larger than the characteristic energy of the local spin fluctuation $T_{\rm K}$.
The formula for the ratio of two interactions is also determined.
\end{abstract}

\maketitle

There has been a long standing issue on strength of spin-orbit coupling since Anderson pointed out its importance in heavy-fermion superconductivity\cite{Anderson84}.
In heavy-fermion systems, the bare spin-orbit coupling is considerably large compared to a renormalized bandwidth $E_{\rm F}^*$\cite{Yamada86,Rice85,Auerbach86}, in which heavy-mass quasiparticle forms Cooper pairs\cite{Kitaoka05}.
Experimentally however, there are several contradictions to this naive expectation.
In the triplet superconductor, UPt$_3$, the so-called ${\bm d}$-vector can rotate freely in application to relatively low magnetic fields\cite{Tou96,Tou98,Joynt02}, which suggests a weak pinning force due to the spin-orbit coupling.
Moreover, the potential candidates of the triplet superconductors, such as UNi$_2$Al$_3$ and URu$_2$Si$_2$, exhibit power-law temperature dependence indicating the presence of line of zeros\cite{Kitaoka05}, which is inconsistent with the group theoretical argument based on the strong spin-orbit coupling\cite{Anderson84,Volovik84,Ueda85,Blount85}.

The key to resolve the inconsistency is a renormalization of the spin-orbit coupling inherent from a formation of the quasiparticle.
Miyake phenomenologically argued this point taking the lowest Kramers doublet in crystalline-electric-field (CEF) as a Wannier basis of the quasiparticle tight-binding model\cite{Miyake85}, where the bare strong spin-orbit coupling has been strictly taken into account.
Provided that the primary quasiparticle band is well isolated from any other bands, residual rotational symmetry-breaking interactions are of the order of $(E_{\rm F}^*/\Delta)^2$, $\Delta$ being the splitting of the bands.
It ensures an approximate conservation of the rotational symmetry in the quasispin (Kramers doublet) space.
Although the scenario seems to be plausible, it is highly non-trivial whether the simple one-body picture properly works for the quasiparticle and the CEF excited states or not, as we recall many-body aspect of the problem.

In this paper, we demonstrate the renormalization of the spin-orbit coupling as well as the quasiparticle formation, and elucidate the criterion for weak residual spin-orbit interaction in the heavy-fermion systems.
For this purpose, we solve the impurity Anderson model with two set of Kramers doublets using Wilson's numerical renormalization-group method\cite{Wilson75,Krishna80,Krishna80a,Sakai89,Costi94}, which is known to be numerically exact for impurity Anderson models.
Viewing a renormalized local interaction as a source of the momentum-dependent effective interaction for superconductivity\cite{Yanase03}, a comparison in strength among local interactions gives a useful threshold for an irrelevancy of the residual spin-orbit interaction.
We assume that the crystal has the inversion symmetry, otherwise the quasispin degeneracy of the band will be lifted.
In this case the spin-orbit coupling competes with the condensation energy rather than the attractive interaction leading to the superconducting state.
It is a separate issue from the present paper\cite{Anderson84,Frigeri04,Sergienko04}.

We begin with the impurity Anderson model with two Kramers doublets,
\begin{equation}
H = \sum_{\gamma\sigma}\left[\sum_{\bm k}\epsilon_{\bm k}c^\dagger_{{\bm k}\gamma\sigma}c_{{\bm k}\gamma\sigma}+V(f^\dagger_{\gamma\sigma}c_{{\bm k}\gamma\sigma}+{\rm h.c.})\right]+H_f,
\end{equation}
where we have neglected ${\bm k}$ and orbital ($\gamma$) dependences in the hybridization and the dispersion of the conduction electrons. We restrict ourselves to the case of infinite Coulomb repulsions among all orbital and spin states for simplicity.
As a simple example for two Kramers doublets, we adopt $|\Gamma_7\pm\rangle=|\pm1/2\rangle$ and $|\Gamma_9\pm\rangle=|\mp3/2\rangle$ in the hexagonal symmetry ($\gamma=7$ or $9$).
Since $|m\rangle$ represents the eigenstate of the $z$-component of the {\it total} angular momentum ${\bm j}$, the bare spin-orbit coupling is strictly taken into account at this stage.
Then, the $f^0$-$f^1$ restricted $f$-electron Hamiltonian reads
\begin{equation}
H_f=\sum_{\sigma}\biggl[ E_f|\Gamma_7\sigma\rangle\langle\Gamma_7\sigma|+(E_f+\Delta)|\Gamma_9\sigma\rangle\langle\Gamma_9\sigma|\biggr],
\end{equation}
where $\Delta$ is the splitting of two doublets, and is the adjustable parameter in the following discussion.
We fix the other parameters as $E_f=-1$ and $W=V^2=0.05$ in unit of the half-width of the conduction band.

\begin{figure}
\includegraphics[width=8.5cm]{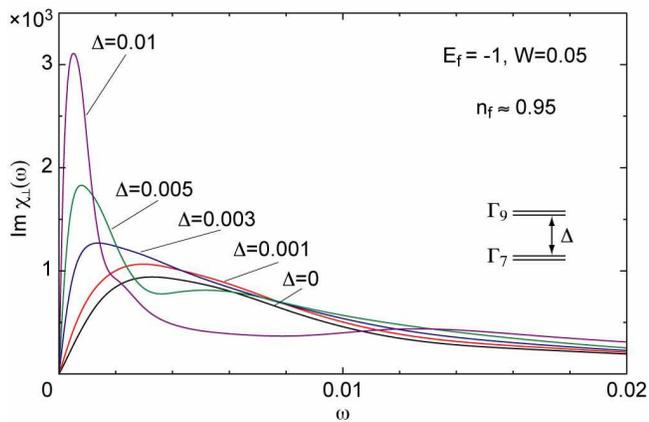}
\caption{The spectral intensity of the dynamical transverse susceptibility for several CEF splittings $\Delta$ at $T=0$. For $\Delta>T_{\rm K}$ ($T_{\rm K}$ is determined by the (lower) peak position), the second broad peak at $\omega\sim\Delta$ is pronounced, otherwise the CEF excitation is obscured by the Kondo resonance scattering.}
\label{chi_t}
\end{figure}

First, let us characterize (local) quasiparticle for several values of $\Delta$.
Figure~\ref{chi_t} shows the spectral intensity of the dynamical transverse susceptibility, ${\rm Im}\chi_\perp(\omega)$ at $T=0$.
As $\Delta$ increases, the quasi-elastic peak becomes sharper with its position going downward.
This is understood by the suppression of spin fluctuations between orbitals, which gives rise to a reduction of the Kondo energy scale $T_{\rm K}$.
Here we have defined $T_{\rm K}$ at which ${\rm Im}\chi_\perp(\omega)$ shows the maximum.
When $\Delta$ exceeds $T_{\rm K}$ (in cases of $\Delta=0.005$, $0.01$), the second broad peak at $\omega\sim\Delta$ corresponding to the CEF excitation is pronounced, otherwise the CEF excitation is obscured by the Kondo resonance scattering\cite{Maekawa85}.

\begin{figure}
\includegraphics[width=8.5cm]{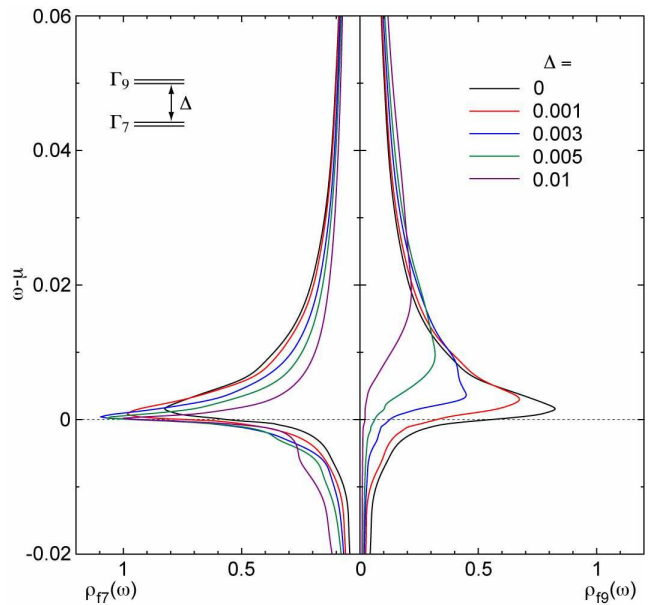}
\caption{The $f$-electron density of states for $\Gamma_7$ (left panel) and $\Gamma_9$ (right panel) orbitals. With increasing $\Delta$, the quasiparticle peak becomes sharper in the lower $\Gamma_7$ orbital, while the peak in the upper $\Gamma_9$ orbital turns into broad incoherent part of the spectrum.}
\label{rho}
\end{figure}

\begin{figure}
\includegraphics[width=8.5cm]{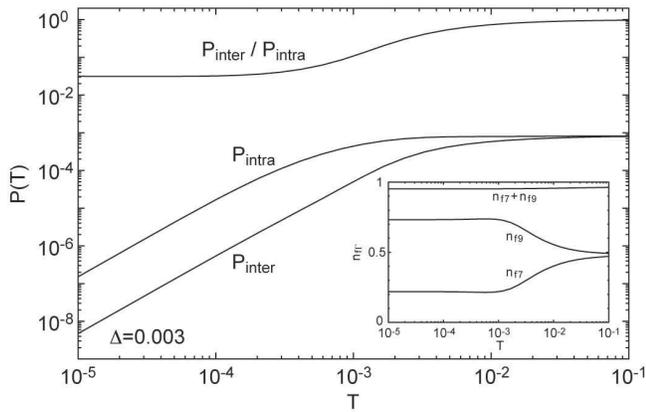}
\caption{Temperature dependence of the effective transition probability of the intra- and the inter-orbital spin-flip operators for $\Delta=0.003$ (see text in detail).
The ratio of two probabilities is unity in the region, $T>\Delta$, while it is considerably small and $T$ independent for $T<T_{\rm K}$, provided that $\Delta>T_{\rm K}$.
The inset shows the thermal average of $f$-electron number in each orbitals.}
\label{mat_t}
\end{figure}

The $f$-electron density of states for each orbitals $\rho_{f\gamma}(\omega)$ is shown in Fig.~\ref{rho}.
Due to the particle-hole asymmetry in the model, the Kondo resonance (local quasiparticle) peak appears slightly above the Fermi energy.
Reflecting the reduction of $T_{\rm K}$ with increase of $\Delta$, the quasiparticle peak becomes sharper in the lower $\Gamma_7$ orbital, while the peak in the upper $\Gamma_9$ orbital turns into the broad incoherent part of the spectrum located at $\omega\sim2\Delta$.
Since with increasing $\Delta$ the integrated spectral weight below the Fermi level increases (decreases) in the $\Gamma_7$ ($\Gamma_9$) orbital, the $f$-electron number changes toward unity (zero) (see also the inset of Fig.~\ref{mat_t}).
Note that the total $f$-electron number is almost independent of $\Delta$, i.e., $n_f\approx 0.95$.
Therefore, we can conclude that the dominant component of the quasiparticle comes from the lower CEF doublet via the local spin fluctuation with conduction electrons, as long as the CEF excitation is clearly observed in the neutron scattering intensity.

Next, we discuss the renormalization of the spin-orbit coupling.
As was mentioned that the bare spin-orbit coupling has already been included in the definition of the CEF state, we consider here two type of residual interactions, i.e., quasispin conserving and non-conserving interactions, respectively.
To quantify the extent of the renormalization, let us define the effective transition probability for an operator $\hat{O}$ as
\begin{equation}
P_O\equiv\sum_{mn}e^{-\beta(E_m+E_n)}|\langle n|\hat{O}|m\rangle|^2/Z^2,\;\;\;Z=\sum_ne^{-\beta E_n}
\end{equation}
where $E_n$ and $|n\rangle$ are the exact eigen energy and its eigenstate, respectively, and $\beta=1/T$ is the inverse temperature.
Then, we consider $\hat{O}_{\rm inter}\equiv f^\dagger_{9\downarrow}f_{7\uparrow}$ for the quasispin non-conserving operator, which is the essential part of the ``spin-orbit'' coupling, and $\hat{O}_{\rm intra}\equiv f^\dagger_{7\uparrow}f_{7\downarrow}$ for the conserving operator.
Note that $P_O$ with $\hat{O}_{\rm intra}$ (we denote it as $P_{\rm intra}$ hereafter) should be equal to $2P_O$ with $\hat{O}=(f^\dagger_{7\uparrow}f_{7\uparrow}-f^\dagger_{7\downarrow}f_{7\downarrow})/2$ due to the Pauli principle.

The temperature dependence of $P_{\rm intra}$ and $P_{\rm inter}$ for $\Delta=0.003$ is shown in Fig.~\ref{mat_t}.
The thermal average of $f$-electron number in each orbitals is also shown in the inset of Fig.~\ref{mat_t}.
As we expect, the orbital polarization in the $f$-electron occupancy is pronounced for $T<\Delta$.
For $T>\Delta$ two probabilities are almost $T$ independent, and have the same magnitude.
In the region $T_{\rm K}<T< \Delta$, $P_{\rm inter}$ begins to decrease rapidly while $P_{\rm intra}$ remains $T$ independent.
With further decrease of $T$, $P_{\rm intra}$ also starts to decrease with the same exponent as that of $P_{\rm inter}$.
Then, the ratio of two probabilities, $r(T)\equiv P_{\rm inter}/P_{\rm intra}$ is unity in the region, $T>\Delta$, while $r(T)\ll 1$ and $T$ independent for $T<T_{\rm K}$, provided that $\Delta>T_{\rm K}$.
In other words, when the quasiparticle is formed below $T_{\rm K}$, and if $T_{\rm K}$ is smaller than the first CEF excitation energy, the quasispin non-conserving interaction is negligible as compared with the quasispin conserving interaction.
Consequently, the pairing interaction for the quasiparticle has the approximate rotational symmetry in the quasispin space.

Now, we discuss more accurate criterion for the irrelevancy of the rotational symmetry-breaking interaction as a function of $T_{\rm K}$ and $\Delta$.
In general, the transition probability near the fixed point is expressed as
\begin{equation}
P_O\sim (T/T_O^*)^{2d_O},
\end{equation}
where $T_O^*$ is the characteristic energy of the operator $\hat{O}$ and $d_{O}$ is its scaling dimension in the scale transformation near the fixed point\cite{Wilson75,Affleck95}.
Near the high-temperature fixed point, the $f$-component in the wavefunction is localized.
Then, both operators, $\hat{O}_{\rm intra}$ and $\hat{O}_{\rm inter}$ have the scaling dimension $d=0$, yielding the $T$ independent behaviors at high temperatures.
For $T<\Delta$, $P_{\rm inter}$ sets in the regime of the low-temperature fixed point, while $P_{\rm intra}$ stays in the regime of the high-temperature fixed point.
Since the $f$ wavefunction of the $\Gamma_9$ orbital is delocalized in the low-temperature regime, the contribution from the impurity site decreases exponentially in the scale transformation.
Indeed, the local hopping such as $\hat{O}_{\rm inter}$ has the Fermi-liquid scaling dimension, $d=1$.
Then, $P_{\rm inter}$ displays the relation $\ln(P_{\rm inter})\sim 2\ln(T/T^*)$ for $T<T^*\sim\Delta$.
For $T<T_{\rm K}$, $P_{\rm intra}$ enters the low-temperature regime as well, showing the similar $T$ dependence as $P_{\rm inter}$ with $T^*\sim T_{\rm K}$ in this case.
In this way we can understand the $T$ dependences obtained in Fig.~\ref{mat_t}.

\begin{figure}
\includegraphics[width=8.5cm]{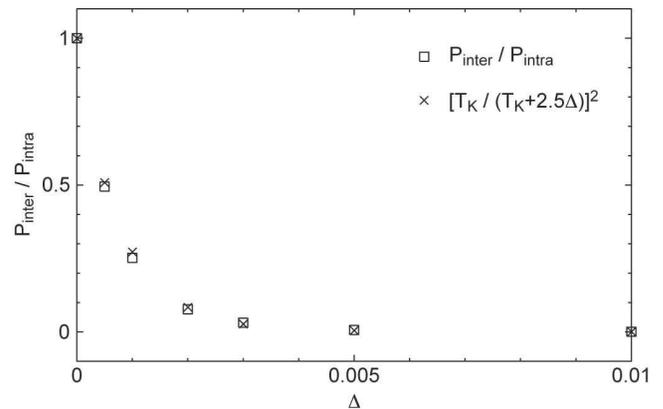}
\caption{The ratio of the intra- and the inter-transition probabilities as a function of the CEF splitting (square). The $\Delta$ dependence of the ratio is well described by the formula, $(T_K/(T_K+\alpha \Delta))^2$ with $\alpha=2.5$ (cross).}
\label{mat_ratio}
\end{figure}

We plot the zero-temperature limit of the probability ratio $r_0=r(T\to0)$ as a function of $\Delta$ in Fig.~\ref{mat_ratio} (square).
From the above discussion, $r_0\sim (T_{\rm intra}^*/T_{\rm inter}^*)^2$ at low temperatures.
If we adopt $T^*_{\rm intra}\sim T_{\rm K}$ and $T^*_{\rm inter}\sim T_{\rm K}+\alpha\Delta$, $\alpha$ being of the order of unity, we have $r_0\sim \bigl[ T_{\rm K}/(T_{\rm K}+\alpha \Delta)\bigr]^2$.
Indeed, the ratio of two probabilities is well described by the formula with $\alpha=2.5$ as shown in Fig.~\ref{mat_ratio} (cross).
It should be noted that $T_{\rm K}$ itself strongly depends on $\Delta$, namely, the increase of $\Delta$ suppresses $T_{\rm K}$.
As a result, the ratio decreases faster than $\Delta^{-2}$.
Recalling the exponential dependence in the critical temperature of the BCS formula, quasispin non-conserving interactions are practically irrelevant to the Cooper-pair formation in the quasiparticle band.

It is interesting to note that a similar situation occurs in Sr$_2$RuO$_4$.
In this case, the bare spin-orbit coupling $\lambda$ rather than the renormalized one appears in the second order, $\lambda^2/E_{\rm F}^*$, because the primary $\gamma$-band for the superconductivity is isolated from the other $\alpha\beta$-bands due to the symmetry reason\cite{Yanase03}.
It can be said that in the heavy-fermion systems, the process of the heavy quasiparticle formation dynamically provides an isolated quasiparticle band similar to the $\gamma$-band in Sr$_2$RuO$_4$.

So far, we have assumed the $f$-electron valency is less than one.
In this case, a characteristic of the quasiparticle in the lattice system and a process of its formation are well understood\cite{Yamada86,Rice85,Auerbach86}.
On the other hand, we have no concrete understanding for a situation of more than two $f$-electron valency.
In particular, the $f^2$ configuration with a singlet CEF ground state, which is believed to be realized in some U-based compounds and Pr-based filled skutterudites\cite{Steglich95,Maple96,Onuki04,Cox98,Aoki05,Maple05}, is remarkable.
This is because the $f$ electrons themselves can release their entropy without hybridizing the conduction electrons\cite{Ikeda97,Kusunose05,Nozieres05}, which is necessary to form the Fermi liquid for systems with the $f^1$ configuration.
Such subjects are currently under extensive investigations\cite{Onuki04,Cox98,Aoki05,Maple05,Ikeda97,Kusunose05,Nozieres05}.

In summary, we have studied the formation of the quasiparticle as well as the renormalization of the residual spin-orbit interaction based on the impurity Anderson model with two set of Kramers doublets.
Using the numerical renormalization-group calculation, we have shown the effective irrelevancy of the rotational symmetry-breaking interaction in the quasispin space for the CEF splitting $\Delta$ larger than the local spin-fluctuation energy $T_{\rm K}$.
As the temperature decreases, the delocalization of the $f$ electron first occurs in the upper CEF state.
Subsequently, the $f$ electron in the lower CEF state delocalizes to form the heavy-mass quasiparticle.
As a result, the inter-orbital interaction, which generates quasispin non-conserving interaction, is smaller than the intra-orbital quasispin conserving interaction by a factor of $(T_{\rm K}/\Delta)^2$.
Since $T_{\rm K}$ has strong $\Delta$ dependence, $(T_{\rm K}/\Delta)^2\ll1$ is easily realized even for small $\Delta$.
Therefore, in most cases the residual spin-orbit coupling is irrelevant for the heavy-fermion superconductors.

The author would like to thank Y. Yanase, Y. Matsuda, H. Tou, K. Miyake and Y. Kuramoto for stimulating discussions.
This work was supported by a Grant-in-Aid for Scientific Research Priority Area ``Skutterudite" of the Ministry of Education, Culture, Sports, Science and Technology, Japan.

\end{document}